\documentclass[1p]{elsarticle}
\usepackage[numbers]{natbib}
\usepackage{amssymb}
\usepackage{delarray}
\usepackage[]{graphicx}
\newenvironment{Cases}{\begin{array}\{{l}.}{\end{array}}

\begin{document}
\begin{frontmatter}
\title{Finite Domain Anomalous Spreading Consistent with First and Second Law}
\author[au1]{P. P. Valk\'o}
\ead{p-valko@tamu.edu}
\author[au2]{X. H. Zhang}
\address{Texas A\&M University}
\address[au1]{Department of Petroleum Engineering}
\address[au2]{Department of Physics}

\begin{abstract}
After reviewing the problematic behavior of some previously suggested finite interval spatial operators of the symmetric Riesz type, we create a wish list leading toward a new spatial operator suitable to use in the space-time fractional differential equation of anomalous diffusion when the transport of material is strictly restricted to a bounded domain. Based on recent studies of wall effects, we introduce a new definition of the spatial operator and illustrate its favorable characteristics. We provide two numerical methods to solve the modified space-time fractional differential equation and show particular results illustrating compliance to our established list of requirements, most important to the conservation principle and the second law of thermodynamics.
\end{abstract}

\end{frontmatter}

\section{Introduction}
\label{sec1}
In practical sense anomalous diffusion can be detected by heavy tails of the resulting density distribution (at a given time) and by the departure from linearly evolving mean-square displacement for an initially concentrated plume \cite{Schneider89,Shlesinger93,Mainardi97}. Such a behavior is well documented for instance for the spreading of contaminants in heterogeneous porous media, where shortcut pathways may be present between two points in space (causing a departure from Fick's law)  and/or particles can be trapped, hindered at various locations (causing memory effects)\cite{Berkowitz95, Benson00a,Benson00b,Metzler00,Metzler04}. With a common term, the behavior may be non-local both in space and time \cite{Zaslavsky02,Chechkin06,Zhang09}.

Continuous time random walks (CTRW) serve as a small-scale conceptual model for describing anomalous diffusion. Of practical interest is the evolution of the density of a cloud of walkers on the macroscopic scale that is ultimately determined by the statistical characteristics of the jump lengths and waiting times on the microscopic level. The so called L\'evy flights are dominated by rare but large jumps and can reproduce the power-law tails of the spatial distributions at a given time. Allowing rare but long waiting times can also lead to marked departure from the scaling law of diffusion. The flux computed on the scale of particle motion, depends on the parameters of the random walk and on the density at the considered time (in the Markovian case) or on the complete density history (in the case of memory effects included). The passage from microscopic to  macroscopic scale is performed by letting the characteristic length and time of the particle motion tend to zero \cite{Montroll65,Cushman91,Barkai02}.  On unbounded domain, the resulting macroscopic behavior is conveniently described by the space-time fractional diffusion equation  \cite{Gorenflo02}. The passage also results in the generalization of Fick's law  \cite{Paradisi01, Neel07}.

\subsection{Brief summary of known results for the unbounded domain case}
The one-dimensional space-time fractional diffusion equation is written as
\begin{eqnarray}
\label{equ1}
\frac{\partial ^\beta}{\partial t^\beta} u(x,t) = a  \frac{\partial ^\alpha}{\partial |x|^\alpha} u(x,t)\;\;\; - \infty < x < \infty, \; t > 0
\end{eqnarray}
where $a$ is positive constant with dimension $[L^\alpha / T^{\beta}]$. (The dimension of $u$ is $[1/L]$ because it is understood as one dimensional density of a countable quantity.)

The fractional time derivative is taken in the Caputo sense \cite{Caputo67}. The notation for the operator $\frac{\partial ^\alpha}{\partial |x|^\alpha} $ was introduced by Saichev and Zaslavsky \cite{Saichev97}. It is understood as the application of the (symmetric) Riesz derivative
$\frac{d ^\alpha}{d |x|^\alpha} $ operator with respect to the space variable $x$.
The Riesz derivative is defined through the Liouville-Weyl fractional derivatives:
\begin{eqnarray}
\frac{d ^\alpha}{d |x|^\alpha} f(x) &=&
\begin{Cases}
-\frac{1}{2 \cos(\pi \alpha /2 )} [  D^\alpha_+ f + D^\alpha_- f ],\;\;\;  0 < \alpha \leq 2,\;\alpha \neq 1\\
-\frac {d}{dx}\textbf{H} (f ; x),\;\;\;\alpha=1\\
- f,\;\;\; \alpha=0
\end{Cases}
\end{eqnarray}
where $D^\alpha_\pm$ are called the left- and right Liouville-Weyl derivatives:
\begin{eqnarray}
\label{LW1}
D^\alpha_+ &=& \frac{1}{\Gamma(m-\alpha)} \frac{d^m}{dx^m} \int_{-\infty}^{x} \frac{f(\xi)d\xi}{(x-\xi)^{\alpha-m+1}},\;\;\;  m = \lceil \alpha \rceil\\
\label{LW2}
D^\alpha_- &=& \frac{(-1)^m}{\Gamma(m-\alpha)} \frac{d^m}{dx^m}  \int_{x}^{\infty} \frac{f(\xi)d\xi} {(\xi-x)^{\alpha-m+1}}.
\end{eqnarray}
and \textbf{H} denotes Hilbert transform. (See a more detailed discussion, for instance, by Chechkin \textit{at al.} \cite{Chechkin08} )

For the unbounded case the following Cauchy problem can be stated: solve  (\ref{equ1}) for a given parameter set $\{ 0<\alpha\leq2,\; 0<\beta\leq1,\; a > 0 \}$ augmented with the initial condition $u(x,0)=f_i(x)$, where $f_i$ is a probability density function. The solution of the Cauchy problem can be obtained by the Laplace-Fourier approach, probably first used in this context by Montroll and Weiss \cite{Montrol73}. The transforms serve double purpose: they provide a better understanding of the operators involved and also lead to the solution for particular cases. In fact, the $\beta$-order Caputo derivative ($0<\beta\leq 1$) is the generalization of the \emph{first derivative} via Laplace transform:
\begin{eqnarray}
 \textbf{L}\left(  \frac{d ^\beta f }{d t^\beta} ; s  \right) = s^{\beta-1}  \left[ s \textbf{L}(f; s) -f(0)\right]
\end{eqnarray}
and the $\alpha$ order Riesz derivative ($0\leq\alpha\leq 2$) is the generalization of the \emph{second derivative} via Fourier transform:
\begin{eqnarray}
 \textbf{F}\left(  \frac{d ^\alpha f }{d |x|^\alpha} ;\omega \right) = -|\omega|^\alpha \textbf{F}(f ;\omega)
\end{eqnarray}

The fundamental solution (spreading of an initial Dirac delta) can be obtained by the Laplace-Fourier method and can be given in terms of well investigated special functions. Though various representations are available, their equivalence has been well established \cite{Mainardi01,Kilbas06} . The remarkable scaling property of the fundamental solution can be stated as:
\begin{eqnarray}
\label{fundsol}
u(x,t) =  (a t)^{-\frac{\beta}{\alpha}}\textit{M}\left(  \frac{|x|}{(a t)^\frac{\beta}{\alpha} } ; \frac{\beta}{\alpha} \right)
\end{eqnarray}
where $\textit{M}$ denotes the Mainardi (or $\textit{M}$) -function given by
\begin{eqnarray}
\textit{M}(\xi; \mu)= \frac{1}{2} \sum_{n=0}^\infty  \frac{\left(-\xi\right)^n}{n! \; \Gamma\left[-\mu n +(1-\mu)\right]}
\end{eqnarray}
(By including the factor $1/2$ we ensure that the integral of (\ref{fundsol}) over the $x$-axis is unity.) While computability of the $\textit{M}$-function is far from trivial, it has been basically resolved. For instance, \textit{Mathematica} can calculate it \emph{with any desired accuracy} from its definition above, provided the $\mu$ parameter is passed as a rational fraction. Some known special cases, see e.g. \cite{Piryatinska05} can be reproduced symbolically with $Mathematica$:

\begin{eqnarray}
\textit{M}(\xi ; 1/2) &=& \frac{1}{2 \sqrt{\pi}}  \exp(-\xi^2 /4)\\
\label{M13}
\textit{M}(\xi ; 1/3) &=& \frac{3^{2/3}}{2} Ai(\frac{\xi}{\sqrt[3]{3}})  \\
\textit{M}(\xi ; 2/3) &=& \frac{1}{2 \; 3^{2/3}} \exp(-\frac{2 \xi^3}{27}) \left[
\sqrt[3]{3} \; \xi \; Ai(\frac{ \xi^2}{3 \; 3^{1/3}}) - 3 Ai'(\frac{ \xi^2}{3 \; 3^{1/3}})\right]
\end{eqnarray}
where $Ai$ stands for the Airy function and  and $Ai'$ for the Airy-prime function.

\subsection{Anomalous spreading on a finite interval}
For Fickian diffusion the finite domain model consistent with the first law (conservation principle) is obtained by requiring zero flux at the two endpoints of the considered interval. This translates to the well-known homogenous Neumann boundary conditions for the traditional $\{\beta=1, \; \alpha=2\}$ diffusion equation. Mapping the same physical requirement to mathematically treatable objects for the space-time fractional partial differential equation has turned out to be extremely challenging.

Numerical experimentation followed two complementary approaches: Monte-Carlo (Langevin) simulation of random walk and finite difference approximation to the solution of a space-time fractional differential equation on bounded domain. The first is easier to conduct and always results in physically meaningful results (including conservation, if reflective walls are applied), but those results are difficult to use in a practical sense. The second approach has grown mature during the years. Here we will rely on the well documented "matrix approach" suite by Podlubny \textit{et al.} \cite{Podlubny09} that provides a general framework to the numerical solution of partial fractional differential equations.

However, the problem is deeper than purely finding an adequate numerical method. In the presence of wall(s) of various properties the spatial operator itself needs to be modified, see e.g. \cite{Chechkin03,Metzler07}.

In this work we are looking for the solution of the Cauchy problem for the fractional partial differential equation
\begin{eqnarray}
\label{problem}
\frac{\partial ^\beta}{\partial t^\beta} u(x,t) = a  \frac{\partial ^\alpha}{\partial_{mod} |x|^\alpha} u(x,t)\;,\;\;\;\; 0 \leq x \leq 1 , \;\;\;\; t > 0
\end{eqnarray}
when $a > 0$  and the initial condition $u(x,0) = f_i(x), \;\;  0 \leq x \leq 1 $ is a probability density function. The fractional time derivative of order ($0<\beta\leq1$) is in the Caputo sense. We added a subscript "mod" to the spatial operator of order ($0<\alpha\leq2$), because  the definitions (\ref{LW1}-\ref{LW2}) require the extension of the $u(x,t)$ function defined on the interval $[0,1]$ to a left function $f_l(x)$ defined on $(-\infty,1]$ (or, strictly speaking, at least on $(-\infty,x]$,) and to a right function $f_r(x)$ defined on $[0,\infty)$. Notice that these auxiliary functions need not be probability distributions. For brevity, we will call the set of choices we make in creating these extensions a "prescription". Various prescriptions will give rise to various finite domain Riesz operators and will ultimately define the characteristics of the solution of (\ref{problem}).

In the following section (\ref{sec2}) we illustrate the problematic behavior of some previously suggested prescriptions and create a "wish list" leading toward a new spatial operator. The subsequent section (\ref{sec3}) introduces the new prescription suitable to treat anomalous spreading on a bounded domain and describes its main characteristics. The (\ref{sec4}) provides two numerical methods to solve the modified space-time fractional differential equation and shows particular results illustrating the key issues. We finish the paper with summary and conclusions.

\section{Finite domain approaches}
\label{sec2}
The mathematically straightforward prescription to create $f_l(x)$ and $f_r(x)$ is padding the function with zero from both sides:
\begin{eqnarray}
\label{prescription 1}
\nonumber
f_l(x)&=& \begin{Cases}
u(x,t),\;\;\;  0 \leq x \leq 1 \\
0,\;\;\; -\infty < x < 0
\end{Cases} \\
f_r(x)&=& \begin{Cases}
u(x,t),\;\;\;  0 \leq x \leq 1 \\
0,\;\;\;  1 <  x <  \infty
\end{Cases}
\end{eqnarray}
With such a prescription, the operator $D^\alpha_+$ will yield the left finite Riemann-Liouville fractional derivative $ _0D^\alpha_x u(x,t)$ and the operator $D^\alpha_-$ will yield the right finite Riemann-Liouville fractional derivative $ _xD^\alpha_1 u(x,t)$, where the integrals are defined already only over the appropriate part of the interval $[0,1]$, see \cite{Podlubny99}. In most numerical calculations published so far such a prescription has been used. It is also the default in the matrix approach. Starting from an $f_i(x)$ probability distribution obeying $f_i(0)=f_i(1)=0$, it seems reasonable to augment the problem with the two Dirichlet boundary conditions $u(0,t)=u(1,t)=0$ and solve it numerically by finite differences.

It is well known however, that the solution with Dirichlet conditions will not satisfy the first law, even in the case of $\beta=1$ and $\alpha=2$. Fig 1. shows the solution of a well documented problem with data $\{\beta=1/2,\;\alpha=3/2,\;a=1,\; f_i(x) = 6 x (1-x)\}$ obtained via the matrix approach suite. (Notice that the actual a parameter passed on should be $a' = \sqrt2$ instead of $a=1$ because in that suite the Riesz derivative is understood without the factor $\cos(\pi \alpha /2 )$.) In the illustration we also show the evolution of the first integral of the spatial density that is the fraction of substance remaining in the finite domain. As it is obvious, material is lost during the process. Replacing the two boundary conditions by "fractional derivative of order $\alpha -1$ equal to zero" condition -- motivated by some interpretation of Fick's law -- does not help either. The problematic behavior of prescription (\ref{prescription 1}) has been repeatedly discussed, for instance, in the groundwater literature \cite{Zhang06}.

Here we show, that no boundary condition can be found to reconcile the non-physical nature of prescription (\ref{prescription 1}). To this end we introduce a small change into the problem specification. Instead of requiring something at the two boundaries, we require that the numerical solution preserve the two important characteristics of the initial distribution: the first integral over $[0,1]$ is equal to unity and it is symmetric. To satisfy these conditions instead of the two Dirichlet boundary conditions we require $\int^1_0 u(x,t) dx = 1$ and $u(0,t)=u(1,t)$. (The second one is obviously necessary, but it turns out to be sufficient as well.) The two conditions are easily passed on to any finite-difference method, in this case to the matrix approach suite \cite{Podlubny09}. Summarized in Fig. 2 are the results for $f_i(x)= 6 x (1-x)$. Forcing the model to satisfy the first law, we lost compliance to the second law. The normalized entropy ($\sum_{i=1}^n{u_i \ln{u_i}}/\ln{\frac{1}{n}}$, where $n$ is the number of spatial mesh points) is \emph{decreasing} with time.

\subsection{Turning to Caputo's idea}
Similar experiences known for practitioners have led to various ideas. For instance, del-Castillo-Negrete \textit{et al.} \cite{Del-Castillo-Negrete08}  suggested to use a modification of the Riesz derivative operator, following the recipe of Caputo being so successful for the time derivative. In our terminology, the prescription (explicitly given here only for $ 1 < \alpha \leq 2$)
\begin{eqnarray}
\label{prescription 2}
\nonumber
f_l(x)&=& \begin{Cases}
u(x,t) -u(0,t)- x \: \left[\frac{\partial}{\partial x}u(x,t)\right]_{\& \; x=0},\;\;\;  0 \leq x \leq 1 \\
0,\;\;\;  -\infty < x < 0
\end{Cases}  \\
f_r(x)&=& \begin{Cases}
u(x,t) -u(1,t) + (1-x) \: \left[\frac{\partial}{\partial x}u(x,t)\right]_{\& \; x=1},\;\;\;  0 \leq x \leq 1 \\
0,\;\;\;  1 <  x  <  \infty
\end{Cases}
\end{eqnarray}
results in the Caputo form of the finite Riesz derivative. (Notice that the prescription introduces a jump discontinuity between the two functions $f_l$ and $f_r$.) Initiating the spreading process from the uniform distribution will leave the initial state at rest, since the right-hand side of (\ref{problem}) will be identically zero. Thus the prescription resolved a contradiction, but now we have to face another one: starting the process from a triangular distribution will also leave the system at rest.

Constructing the flux expression (Fick's law) directly with the Caputo derivative, Zhang \textit{et al.} \cite{Zhang07} introduced another variant without explicit use of the first spatial derivative. In our notation, the prescription will take the form:
\begin{eqnarray}
\label{prescription Zhang}
\nonumber
f_l(x)&=& \begin{Cases}
u(x,t) -u(0,t),\;\;\;  0 \leq x \leq 1 \\
0,\;\;\;  -\infty < x < 0
\end{Cases} \\
f_r(x)&=& \begin{Cases}
u(x,t) -u(1,t),\;\;\;  0 \leq x \leq 1 \\
0,\;\;\;  1 <  x  <  \infty
\end{Cases}
\end{eqnarray}
The advantage of prescription (\ref{prescription Zhang}) is that an initial triangular distribution will not be a steady-state solution of problem (\ref{problem}) any more. However, it is still easy to find an initial condition (e.g. a box function non-zero only over a part of the [0;1] interval)  that will also be a steady-state solution -- in contrast to physical intuition. Fig 3 illustrates the problematic behavior of  prescriptions (\ref{prescription 1}), (\ref{prescription 2}), and (\ref{prescription Zhang}).

\subsection{Summary of desired characteristics for a finite domain model}
From the introductory numerical experiments we glean a wish list. Sought is a macroscopic model in the form of fractional partial differential equation (\ref{problem}) with the "hard" requirements (1-5) and "soft" ones (6-10):
\begin{enumerate}[i)]
\item If the initial state is a probability distribution (non-negative and with unit area under the curve), this property should be preserved for any time;
\item If the initial probability distribution is symmetric around $x=0.5$, this property should be preserved for any time;
\item The only stable steady state should be the uniform distribution;
\item For any non-uniform initial distribution, the entropy should monotonically increase with time;
\item For integer orders, the model should reduce to known results;
\item Starting from a Dirac delta distribution, the solution should follow the unbounded fundamental solution for short times;
\item We still want to preserve the deep correspondence with Caputo fractional derivative in time and LW fractional derivative in space (allowing, however, some liberty in the selection of the $f_l$ and $f_r$ functions);
\item Motivation should stem from a microscopic CTRW concept;
\item It is desirable to have analytic solution for special cases;
\item It is desirable to have a numerical solution method within the general framework of discrete fractional calculus \cite{Podlubny09} .
\end{enumerate}
Notice that \emph{item i)}  is sometimes stated as the probability preserving property, or the conservation principle. In this work we take the liberty to refer to it as "first law". \emph{Item ii)}  expresses the invariance with respect to directing the coordinate axis, that is we consider only the symmetric Riesz derivative. \emph{Items iii-iv}  are obviously related to the second law (of thermodynamics). In some applications (see e.g. stock prices, \cite{Scalas06}) the listed requirements can be relaxed, but for description of spreading of material they are obviously necessary.

\section{The effects of walls}
\label{sec3}
\subsection{A prescription by Krepysheva \textit{et al.}}
Our starting point is the work of Krepysheva \textit{et al.} \cite{Krepysheva06a} who visualized a "reflecting" wall at location $x=0$ and showed that, due to its non-local character, the kernel of the fractional space derivative has to be modified. The rule for the hopping particle was that if its jump interacts with the wall, it would continue to move in the mirror direction, preserving the overall length of the "initially intended" jump. In the macroscopic limit, the modified Riesz kernel turned out to be markedly different from the standard one based on finite interval left and right Riemann-Luiville derivatives. In our terminology, the works Krepysheva \cite{Krepysheva06a,Krepysheva06b} derived the following specific prescription: Starting from an $u(x,t)$ function available over the non-negative $x$-axis, construct:
\begin{eqnarray}
\label{prescription 3}
\nonumber
f_l(x)&=& \begin{Cases}
u(x,t),\;\;\;  0 \leq x \leq 1 \\
u(-x,t),\;\;\;  -\infty < x < 0 \\
\end{Cases} \\
f_r(x)&=&u(x,t), \;\;\;  0 \leq x < \infty
\end{eqnarray}
Calvo \textit{at al.} \cite{Calvo07}  have developed the idea further, for a rather specific geometric situation, when the random walk is along the perimeter of a circle. Building on these results, van Milligen \textit{at al.} \cite{vanMilligen08} recently suggested a modification of the spatial operator for our problem (\ref{problem}) that involves the Hurwitz zeta function. Another extension of the idea -- based on the so-called Kolwankar-Gangal derivative -- was proposed by N\'{e}el \textit{et al.} \cite{Neel07}. Recently, Zoia \textit{et al.} \cite{Zoia07} also discussed the effect of walls, although not from a first law point of view.

\subsection{A new prescription}
This work suggests another turn in the development for the isolating two-wall case. We recall that the hydrodynamic limit procedure makes use of the fact that in general, measurements correspond to time and length scales much larger than those of particle motions. In our opinion, it follows that even the "extremely long" jumps of a particle must be shorter than the domain size. Our main idea is therefore to limit the jump size to one, and hence allow zero or one particle-wall interaction, but never more than one. This suggests a new prescription: Starting from an $u(x,t)$ function available on $0 \leq x \leq 1$, construct
\begin{eqnarray}
\nonumber
f_l(\xi) &=& \begin{Cases}
u(\xi,t),\;\;\;   0 \leq \xi \leq x \\
u(-\xi,t),\;\;\;  x-1 \leq \xi \leq 0 \\
0,\;\;\; -\infty < \xi < 0
\end{Cases}
\\
\label{prescription new}
f_r(\xi)&=& \begin{Cases}
u(\xi,t),\;\;\;  x \leq \xi \leq 1\\
u(2-\xi,t),\;\;\; 1 < \xi \leq 1 +x\\
0,\;\;\; 1+x < \xi < \infty
\end{Cases}
\end{eqnarray}
The total support of both $f_l$ and $f_r$ is always two units long but it moves with the location of $x$. The  $f_l$ and $f_r$ functions combined contain every function value from the original $u(x,t)$ twice (one corresponding to direct jump and the other bumped from the wall.) Fig. 4 shows the construction of the extensions $f_l$ and $f_r$ for two specific functions and a \emph{specific location} $x=1/3$.

\subsection{The modified Riemann-Luiville-Riesz derivative}
For comparison purposes, we can cast prescription (\ref{prescription new}) into a more familiar form, using the concept of modified left and right  finite-interval Riemann-Luiville derivatives:
\begin{eqnarray}
_{x-1}D^\alpha_{x,mod} f &=& \frac{1}{\Gamma(m-\alpha)} \frac{d^m}{dx^m} \int_{x-1}^{x} \frac{f_{mod}(\xi)d\xi}{(x-\xi)^{\alpha-m+1}}, \;\;\;  m = \lceil \alpha \rceil\\
_{x}D^\alpha_{x+1,mod} f &=& \frac{(-1)^m}{\Gamma(m-\alpha)} \frac{d^m}{dx^m}  \int_{x}^{x+1} \frac{f_{mod}(\xi)d\xi} {(\xi-x)^{\alpha-m+1}}.
\end{eqnarray}
where
\begin{eqnarray}
f_{mod}(\xi) &=& f(2-\xi) \Pi \left(\xi-3/2\right)+f(\xi) \Pi \left(\xi-1/2\right)+
f(-\xi) \Pi \left(\xi+1/2\right) \;\;\;\;\;\;\;\;\;\;
\end{eqnarray}
with the Heaviside box function defined as $\Pi \left(\xi\right)=1$ for $|\xi| \leq 1/2$ and zero otherwise. These finite Riemann-Luiville derivatives are based on an interval of total length 2, always centered at the location $x$ where we are interested in the derivative. Therefore, for non-integer $\alpha$ the modified finite-interval Riemann-Luiville-Riesz derivative takes the form
\begin{eqnarray}
\label{modRLRiesz}
\frac{d ^\alpha}{d_{mod} |x|^\alpha} f(x)  =
-\frac{1}{2 \cos(\pi \alpha /2 )} [ _{x-1}D^\alpha_{x,mod} f +\; _xD^\alpha_{x+1,mod} f]
\end{eqnarray}
Definition (\ref{modRLRiesz}) and prescription (\ref{prescription new}) are equivalent.

It is illuminating to compare the regularly used spatial operator (\ref{prescription 1}) and the modified one (\ref{modRLRiesz}) for some simple functions defined over $[0;1]$. Fig 5 shows the comparison for $f(x)=x-0.5$, $f(x)=(x-0.5)^2$, and $f(x)=\sin(\pi x)$. Somewhat disappointingly, the modified Riesz derivative (\ref{modRLRiesz}) does not eliminate singularity at the end points of the interval for these functions and -- by and large -- behaves similarly to the commonly used definition (\ref{prescription 1}). However, for one function family investigated in the next sub-section, the difference is dramatic.

\subsection{Eigenfunctions and eigenvectors}
Of particular interest is the the family of functions $\cos( j \pi x)$,  $j=0,1, \ldots $. Also shown in Fig. 5 the comparison for $f(x)=\cos(3 \pi x)$. We see that prescription (\ref{prescription 1}) leads to singular behavior at the endpoints -- as usual. On the other hand,  (\ref{modRLRiesz}) is not only non-singular, but it \emph{almost} coincides with the Fourier-Riesz derivative of the entire $\cos(3 \pi x)$ function over the interval $[0,1]$, differing only in a constant factor $1.01328\ldots$. Using $Mathematica$ one can show that
\begin{eqnarray}
\label{eigenfunctions}
\frac{d ^\alpha}{d_{mod} |x|^\alpha} \cos( j \pi x) &=& c_{\alpha,j} \cos( j \pi x) \;\;\; j=0,1,\ldots
\end{eqnarray}
where $c_{\alpha,j}$ is constant. In other words, $\cos( j \pi x)$ is an eigenfunction of the  operator (\ref{modRLRiesz}) with eigenvalue $c_{\alpha,j}$. Moreover, $c_{\alpha,0}=0$, implying that the uniform distribution has a modified Riesz derivative equal to zero. This property will ensure that the uniform distribution will be a steady state solution of (\ref{problem}). We managed to obtain closed form expression for specific $\alpha$ parameters (with repeated help from $Mathematica$) as follows:
\begin{eqnarray}
\label{eigenvalues}
c_{2,j} &=& - {(j \pi)}^2, \;\;\; j=1,2, \ldots \\
c_{3/2,j} &=& -2 {(j \pi)}^{3/2}  C_F(\sqrt{2 j}), \;\;\; j=1,2, \ldots \\
c_{1/2,j} &=& -2 \sqrt{j \pi }  S_F(\sqrt{2 j} ), \;\;\; j=1,2, \ldots
\end{eqnarray}
where $C_F$ and $S_F$ denote the $\cos$ and $\sin$ Fresnel integrals, respectively. For other $\alpha$ values we could not obtain an explicit expression, but could still develop a simple code in $Mathematica$ that calculates the eigenvalue \emph{with any required number of digit} accuracy for an $\alpha$ given as a rational fraction (see Appendix).

\section{The solution of the space-time fractional differential equation on the interval [0,1]}
\label{sec4}
We introduce two approaches. The first one uses Laplace transform in time and collocation in space. The second one is based on finite differences.

\subsection{The Laplace Transform -- Collocation method: an example}
We illustrate this method on a simple example: $\{\beta = 1/2,\;\alpha = 3/2,\; a = 1,\; f_i(x)=1 - \cos(2 \pi x)\}$. Using 6 collocation points: $\textbf{x}_c = \{0, \frac15,\frac25,\frac35,\frac45,1\}$ we seek the Laplace transform of the solution at an arbitrary point, $x$. Introducing the vector notation
\begin{eqnarray}
\textbf{v}(s)&=&\{ c_0(s),c_1(s),c_2(s),c_3(s),c_4(s),c_5(s)\} \\
\textbf{g}(x)&=&\{1,\cos (\pi  x),\cos (2 \pi  x),\cos (3 \pi  x),\cos (4 \pi  x),\cos (5 \pi  x)\}\\
\nonumber
\textbf{e}&=&
\{
0,-2 \pi^{3/2} C_F(\sqrt{2}),-4 \sqrt{2} \pi^{3/2} C_F(2),-6 \sqrt{3} \pi ^{3/2} C_F(\sqrt{6}), \\
&&-16\pi^{3/2} C_F(2\sqrt{2}), -10\sqrt {5}\pi^{3/2} C_F(\sqrt{10} ) \}
\end{eqnarray}
we represent the Laplace space solution at $x$ as
\begin{equation}
\label{Uxs}
U(x,s) =\textbf{v}(s) . \textbf{g}(x)
\end{equation}
Then its modified Riesz derivative of order $3/2$ takes the form
\begin{equation}
\frac{\partial ^{3/2}}{\partial_{mod} |x|^{3/2}} U(x,s) =  \textbf{v}(s).\left[ \textbf{e} \, \textbf{g}(x)\right]
\end{equation}
(with component by component multiplication between $\textbf{e}$ and $\textbf{g}$) and its Caputo derivative of order $1/2$ is written as
\begin{equation}
\textbf{L} \left(  \frac{\partial ^{1/2}}{\partial t^{1/2}} u(x,t),\;s \right)   = s^{-1/2} \left[s \left( \textbf{v}(s) . \textbf{g}(x)\right)  - f_i(x)\right]
\end{equation}
Writing the partial differential equation
\begin{equation}
\label{collocation}
s^{-1/2} \left[s \left( \textbf{v}(s) . \textbf{g}(x)\right)  - f_i(x)\right] \;-\;  \textbf{v}(s).\left[ \textbf{e} \, \textbf{g}(x)\right] = 0
\end{equation}
at the 6 collocation points $\textbf{x}_c$, we obtain a system of linear equations in the 6 unknown variables, $\textbf{v}(s)$. The solution turns out to be
\begin{equation}
\textbf{v}(s)=\{\frac{1}{s},0,-\frac{1}{4 \sqrt{2} \pi ^{3/2} C_F(2) \sqrt{s}+s},0,0,0 \}
\end{equation}
and hence we obtain
\begin{equation}
U(x,s) = \frac{1}{s}-\frac{\cos (2 \pi  x)}{4 \sqrt{2} \pi ^{3/2} C_F(2) \sqrt{s}+s}
\end{equation}
This can be inverted on $s$ resulting in
\begin{equation}
\label{inverted}
u(x,t) =  1 + \exp\left({32 \pi ^3 C_F(2)^2 t}\right)  \left[ \textrm{erf} \left(4 \sqrt{2} \pi ^{3/2} C_F(2) \sqrt{t} \right)-1 \right]\cos (2 \pi  x)
\end{equation}
Increasing the number of collocation points does not change the solution (\ref{inverted}), it is already exact.

\subsection{The fundamental solution for $\beta = 1/2,\;\alpha = 3/2$}
We can repeat the above procedure for $f_i(x)=\delta(x-\frac12) = 1-2 \cos(2 \pi  x) + 2 \cos(4 \pi  x)-2 \cos(6 \pi  x)+ \; \ldots$ and obtain the fundamental solution:
\begin{eqnarray}
\nonumber
u(x,t)&=&
1+2 \exp\left({32 \pi ^3 C_F(2)^2 t}\right) \cos (2 \pi  x) \left[\textrm{erf}\left(4 \sqrt{2} \pi ^{3/2} C_F(2) \sqrt{t}\right)-1\right]-\\
\nonumber
&&2 \exp\left({256 \pi ^3 C_F(2 \sqrt{2})^2 t}\right) \cos (4 \pi  x) \left[\textrm{erf}\left(16 \pi ^{3/2} C_F(2 \sqrt{2}) \sqrt{t}\right)-1\right]+\\\nonumber
&&2 \exp\left({864 \pi ^3 C_F(2 \sqrt{3})^2 t}\right) \cos (6 \pi  x) \left[\textrm{erf}\left(12 \sqrt{6} \pi ^{3/2} C_F(2 \sqrt{3}) \sqrt{t}\right)-1\right]-\\
\label{fundsol3}
&&\ldots
\end{eqnarray}
Looking back at our wish list, we would like to check \emph{item vi)} requiring the correspondence of the unbounded and bounded solutions at early times. The analytic solution of the problem stated on unbounded domain would take the form:
\begin{eqnarray}
\label{fundsol2}
u_\infty(x,t) =  t^{-\frac13}\textit{M}\left(  |x|\, t^{-1/3}  ; 1/3 \right)
\end{eqnarray}
where $\textit{M}(\xi ; 1/3)$ is given by (\ref{M13}). Fig. 6 compares, at an early time, $t=10^{-3}$, the new fundamental solution (\ref{fundsol3}) computed with 50 terms and the well-know unbounded solution (\ref{fundsol2}). We see that the effect of the boundary has just started to show. (Unfortunately, the numerical evaluation of the fundamental solution is not trivial -- even with $Mathematica$ -- because of the extremely large exponents and hence we could not increase the number of terms to get rid of the small oscillations of the curve.)

Pursuing the "analytic" solution further has other drawbacks too. The fundamental solution would contain Mittag-Leffler functions (in addition to the eigenvalues,  $c_{\alpha,j}$  and eigenfunctions $\cos(j\: \pi \:x)$ ) in the case of a general $\alpha$ and  ${\beta}$. Moreover, any other (non Dirac delta) initial condition would necessitate further numerical convolution.

One can, however, easily construct the system of linear equations for a given set of $\alpha$, $\beta$, $a$ and $f_i(.)$ at any selected value of the Laplace variable $s$, solve the system numerically by Gauss elimination and substitute $\textbf{v}(s)$ into (\ref{Uxs}). Therefore, we have a way to calculate (at any specified $x$) the Laplace transform of the solution numerically, and hence we can use a numerical inversion technique \cite{Valko04}. The procedure is robust, if care is taken to do the Gauss elimination with multiple precision -- large enough with respect to the number of terms in the GWR algorithm \cite{Abate04}. We will call the method LT--collocation with numerical inversion. A $Mathematica$ realization of the algorithm is provided in the Appendix.

We emphasize that the procedure does not require any boundary conditions, rather $x=0$ and $x=1$ are included in the set of collocation points. (The physical "boundary conditions" are taken care of within the spatial operator itself.)

Shown in Fig. 7 is the summary of the results of the procedure for  $\{\beta = 1/2,\;\alpha = 3/2,\; a = 1,\; f_i(x)=\delta (x-\frac12)\}$, using $51$ collocation points. Since the LT--collocation method with numerical inversion cannot be started from the Dirac delta "function", we pass on the solution of the unbounded problem  (\ref{fundsol2}) at a very early time, $t = 0.00001$ as a new "initial" condition. Then we do the numerical inversion for $t-0.00001$ where $t$ is the time we are interested in. The "fraction of substance still in the domain" is not shown in Fig. 7, because at the "initial" state it is unity (for all practical purposes) and hence it remains unity during all times.

Next we compare traditional $\{\beta = 1,\;\alpha = 2,\; a = 1\}$ and space-time fractional spreading  $\{\beta = 1/2,\;\alpha = 3/2,\; a = 1\}$ starting from a non-symmetric initial distribution $f_i(x) = 12 x(1-x)^2$, as illustrated by Figs 8 and 9.  The number of collocation points is kept at $51$.  We find that the spreading is initially faster for the space-time fractional case, but at late time traditional dispersion becomes faster.  There is also a remarkable difference in the "overshoot" of the density at $x=0$ -- at least for the studied time points. (We note, however, that such comparisons are of limited value, because the parameter $a$ is not the same in the two cases, in spite of looking the same.)

\subsection{Solution by the method of finite differences}
While the LT--collocation with numerical inversion works effectively, it is still illuminating to solve the problem within the framework of discrete fractional calculus. The form of the modified Riesz operator (\ref{modRLRiesz}) suggests that a relatively small modification to the established matrix approach will suffice. Indeed, one has to use prescription (\ref{prescription new}) to pad the list of available function values and add a small correction to assure that for a constant function the modified Riesz derivatives yield zero.

We introduced this modification into the matrix approach suite (see Appendix for some details). When working with the suite, we do not use the concept of "mathematical boundary conditions" at all, rather we write the equations for the endpoints $x=0$ and $x=1$ as well. Therefore, the total number of equations remains the same as the number of unknowns. Shown in Fig. 10 is the summary of results for our previous example $\{\beta = 1/2,\;\alpha = 3/2,\; a = 1\}$ (that is $a' = \sqrt2$) when the initial condition is $f_i(x) = 12 x(1-x)^2$ and the finite difference step sizes are $\Delta x=0.02$ and  $\Delta t=0.01$. (Notice that the number of unknowns in the matrix approach was 5000 and the system matrix had $2.5$ million elements, hence we were limited by computer memory.) Regarding accuracy, the results are still behind the ones depicted in Fig. 9, but the overall correspondence is remarkable. In particular, the fraction of substance still in the domain has less than 2 \% error. Not only the modified Riesz derivative is "probability preserving" but also is our finite difference representation of it.

\section{Summary and conclusions}
We have introduced a new version of the finite Riesz derivative. The new spatial operator -- combined with the Caputo derivative in time -- results in a space-time fractional differential equation that is well suited to describe anomalous spreading of substance in a finite domain. The space-time fractional differential equation satisfies our postulated requirements. If the initial state is a probability distribution (non-negative and with unit area under the curve), this property is preserved for any later time; if the initial probability density is symmetric around $x=0.5$, this property is also preserved. We could not prove rigorously that the only stable steady state is the uniform distribution, but all specific analytic formulaes and numerical examples indicated so. In all our examples, starting from a non-uniform initial distribution, the entropy monotonically increased with time. For integer orders, the model reproduces the known results of Fickian-Markovian diffusion over a finite domain. Starting from a Dirac delta distribution, the solution follows the unbounded fundamental solution of the space-time fractional differential equation for short times. We could preserve the deep correspondence with Liouville-Weyl fractional derivative in space by creating the appropriate prescription, however we are not sure that this is the only (or even the best) way to do it. The new approach arose from a microscopic CTRW concept and we could manage to provide analytic solution for special cases. While our preferred solution method is the LT--collocation with numerical inversion, we could also extend the matrix approach and hence fit our operator into the mainstream framework of discrete fractional calculus. In this work we focused on "pure" diffusion but we do not envisage any difficulty in considering simultaneous advection or other external potential field.

\section{Appendix}
Calculations were done in $Mathematica$. Here we show some code snippets and results in order to ease reproduction of our results.

\noindent The code snippet for the LT-collocation method with numerical inversion is the following:\newline

\[
\begin{array}{lll}
\textsf{vari}[\textsf{s$\_$}]&=&\textsf{Table}[c[j][s],\{j,0,\textsf{nm}\}];\\
\textsf{cosi}[\textsf{x$\_$}]&=&\textsf{Table}[\textsf{Cos}[ (j \pi )\textsf{  }x],\{j,0,\textsf{nm}\}];\\
\textsf{symi}[\textsf{x$\_$}]&=&\textsf{Table}[\textsf{scos}[\alpha ][j][x],\{j,0,\textsf{nm}\}];\\
\textsf{equl}[\textsf{s$\_$}][\textsf{x$\_$}]&=&s{}^{\wedge} {(\textsf{$\beta $}-1)} \; (s\textsf{  }\textsf{vari}[s].\textsf{cosi}[x]\textsf{  }-\textsf{fi}[x]) - a \; \textsf{vari}[s].\textsf{symi}[x] ;\\
\textsf{xc}&=&N[\textsf{Range}[0,1,1/\textsf{nm}],\textsf{maxprec}];\\
\{\textsf{mb}[\textsf{s$\_$}],\textsf{As}[\textsf{s$\_$}]\}&=&\textsf{Map}[\textsf{Normal},
\textsf{CoefficientArrays}[\textsf{equl}[s][\textsf{xc}],\textsf{vari}[s]]];\\
\textsf{sols}[\textsf{s$\_$}]&:=&\textsf{sols}[s]=\textsf{LinearSolve}[\textsf{As}[s],-\textsf{mb}[s]];\\
\textsf{Uxs}[\textsf{x$\_$}][\textsf{s$\_$}]&:=&\textsf{sols}[s].\textsf{cosi}[x];\\
\textsf{uxt}[\textsf{M$\_$}][\textsf{x$\_$}][\textsf{t$\_$}]&:=&\textsf{GWR}[\textsf{Uxs}[x],t,\textsf{M}];\\
\end{array}
\]

\noindent The user has to provide $\beta, \alpha, a$, the function $fi[x\_]$ in addition to the integer $nm$ (the number of collocation points minus one.) The number of terms in the GWR algorithm $M$ \cite{Abate04} and the maximum used precision $maxprec$ (we used 200) are also required. Once defined, the $\textsf{uxt}[\textsf{M}][\textsf{x}][\textsf{t}]$ function can be used to calculate the solution at a specific $x$ and $t$. The $\textsf{scos}[\alpha ][j][x]$ expression should evaluate to
$-2 j^{3/2} \pi^{3/2} FresnelC[\sqrt{2 j}] Cos[j \pi x]$ when $\alpha =3/2$. In general, it will be the product of an eigenvalue $cei[\alpha][j]$ and the appropriate  $Cos[j \pi x]$. While no general formula is currently available for the eigenvalue, we can calculate it by the following code snippet for a given rational fraction $\alpha$ and positive integer $j$:\newline

\[
\begin{array}{lcl}
\textsf{eug}=\textsf{EulerGamma};\\
h[j][x\_]=\textsf{Cos}[j \: \pi x];
\\
\textsf{cei}[\alpha][j]=N[ \textsf{symRmod}[h[j],\textsf{eug},\alpha]/\textsf{h}[j][\textsf{eug}], \textsf{maxprec}];
\end{array}
\]
where

\[
\textsf{symRmod}[f\_,x\_,\alpha\_]:=-\frac{1}{2 Cos[\alpha \pi /2]} (\textsf{lRmod}[f,x,\alpha]+
\textsf{rRmod}[f,x,\alpha]);
\]

\noindent In the above code the left modified Rieman-Luiville-Riesz derivative is calculated from\newline

\[
\begin{array}{l}
\textsf{lRmod}[\textsf{f$\_$},\textsf{t$\_$},\alpha \_] := \textsf{Module}[\{\textsf{fext},m=\textsf{Ceiling}[\alpha ]\},
\\
\textsf{fext}[\textsf{x$\_$}]=f[x]\;\textsf{UnitBox}[x-1/2]+f[2-x]\;\textsf{UnitBox}[x-3/2]+
f[-x]\;\textsf{UnitBox}[x+1/2];\;\;\;\;
\\
\textsf{If} [ \textsf{IntegerQ}[\alpha ],\;
D[f[t],\{t,\alpha \}],
\\
\frac{1}{
\textsf{Gamma[m - $\alpha$ ]}
}\;
D[(\textsf{Integrate}[\frac{\textsf{fext}[\tau ]}{(\textsf{eug} -\tau ){}^{\wedge}(\alpha +1-m)},
\{\tau ,\textsf{eug}-1,\textsf{eug}\}])\textsf{/.}\textsf{eug} \rightarrow t,
\{t,m \}]]]
\end{array}
\]
and the right modified finite-interval Rieman-Luiville-Riesz derivative is calculated from
\[
\begin{array}{l}
\textsf{rRmod}[\textsf{f$\_$},\textsf{t$\_$},\alpha \_] := \textsf{Module}[\{\textsf{fext},m=\textsf{Ceiling}[\alpha ]\},
\\
\textsf{fext}[\textsf{x$\_$}]=f[x]\;\textsf{UnitBox}[x-1/2]+f[2-x]\;\textsf{UnitBox}[x-3/2]+
f[-x]\;\textsf{UnitBox}[x+1/2];\;\;\;\;
\\
\textsf{If} [ \textsf{IntegerQ}[\alpha ],\;
D[f[t],\{t,\alpha \}],
\\
\frac{(-1)^m}{
\textsf{Gamma[m - $\alpha$ ]}
}\;
D[(\textsf{Integrate}[\frac{\textsf{fext}[\tau ]}{(\tau-\textsf{eug} ){}^{\wedge}(\alpha +1-m)},
\{\tau ,\textsf{eug},\textsf{eug}+1 \}])\textsf{/.}\textsf{eug} \rightarrow t,
\{t,m \}]]]
\end{array}
\]
(The extensive use of "EulerGamma" is somewhat arbitrary but proved useful in the context of the current version -- v7.0 -- of $Mathematica$ \cite{MMA}.)

Without going into details of the derivation, here we illustrate the concept of extending the matrix approach suite of Podlubny  \textit{et al.} \cite{Podlubny09}. For  $1<\alpha\leq2$, the current symmetric Riesz function of the suit (called $ransym$) would hypothetically provide the following array $SR$ of symmetric Riesz derivatives at points $\{0,1h,2h,3h,4h\}$ where the function values $\{y_0,y_1,y_2,y_3,y_4\}$ are known:

\[
\begin{array}{lll}
SR_0&=&
 2^{-1+2 \alpha } (1-\alpha ) y_0+\\&&2^{-1+2 \alpha } y_1 \\
SR_1&=& \frac{1}{3} 4^{-2+\alpha } (24+12 (-1+\alpha ) \alpha ) y_0-\\&&4^{\alpha } \alpha  y_1+\\
&&
\frac{1}{3} 4^{-2+\alpha } (24+12 (-1+\alpha ) \alpha ) y_2-\\&&\frac{1}{3} 4^{-1+\alpha } (-2+\alpha ) (-1+\alpha ) \alpha  y_3+\\
&&
\frac{1}{3} 4^{-2+\alpha } (-3+\alpha ) (-2+\alpha ) (-1+\alpha ) \alpha  y_4 \\
SR_2&=& \frac{1}{3} 4^{-1+\alpha } (2-\alpha ) (-1+\alpha ) \alpha  y_0+\\&&\frac{1}{3} 4^{-1+\alpha } (6+3 (-1+\alpha ) \alpha ) y_1-\\
&&4^{\alpha } \alpha  y_2+\\&&
\frac{1}{3} 4^{-1+\alpha } (6+3 (-1+\alpha ) \alpha ) y_3+\\
&&\frac{1}{3} 4^{-1+\alpha } (2-\alpha ) (-1+\alpha ) \alpha  y_4 \\
SR_3&=& \frac{1}{3} 4^{-2+\alpha } \alpha  \big(-6+11 \alpha -6 \alpha ^2+\alpha ^3\big) y_0-\\&&\frac{1}{3} 4^{-1+\alpha } \alpha  \big(2-3 \alpha +\alpha ^2\big) y_1+\\
&&4^{-1+\alpha } \big(2-\alpha +\alpha ^2\big) y_2-\\&&
4^{\alpha } \alpha  y_3+\\
&&4^{-1+\alpha } \big(2-\alpha +\alpha ^2\big) y_4 \\
SR_4&=& 2^{-1+2 \alpha } y_3+\\&&2^{-1+2 \alpha } (1-\alpha ) y_4\\
\end{array}
\]

A symmetric Riesz function modified according to prescription (\ref{prescription new}) would hypothetically provide the following array $SRM$ for the same input:

\[
\begin{array}{lll}
SRM_0&=&
 -2^{2 \alpha } \alpha  y_0+\\&&2^{-1+2 \alpha } \big(2-\alpha +\alpha ^2\big) y_1-\\&&
 \frac{1}{3} 2^{-1+2 \alpha } \alpha  \big(2-3 \alpha +\alpha ^2\big) y_2+\\&&\frac{1}{3} 2^{-3+2 \alpha } (-3+\alpha ) \alpha  \big(2-3 \alpha +\alpha ^2\big) y_3+\\&&
 \frac{1}{3} 2^{-3+2 \alpha } (4-\alpha ) (-3+\alpha ) \big(2-3 \alpha +\alpha ^2\big) y_4 \\
SRM_0&=&
 4^{-1+\alpha } \big(2-\alpha +\alpha ^2\big) y_0-\\&&\frac{1}{3} 4^{-1+\alpha } \alpha  \big(14-3 \alpha +\alpha ^2\big) y_1+\\&&
 \frac{1}{3} 4^{-2+\alpha } \big(24-18 \alpha +23 \alpha ^2-6 \alpha ^3+\alpha ^4\big) y_2+\\&&\frac{1}{3} 4^{-2+\alpha } \big(-48+92 \alpha -58 \alpha ^2+16 \alpha ^3-2 \alpha ^4\big) y_3+\\&&
 \frac{1}{3} 4^{-2+\alpha } \big(-6 \alpha +11 \alpha ^2-6 \alpha ^3+\alpha ^4\big) y_4 \\
SRM_0&=&
 -\frac{1}{3} 4^{-1+\alpha } \alpha  \big(2-3 \alpha +\alpha ^2\big) y_0+\\&&\frac{1}{3} 4^{-2+\alpha } \big(24-18 \alpha +23 \alpha ^2-6 \alpha ^3+\alpha ^4\big) y_1+\\&&
 \frac{1}{3} 4^{-2+\alpha } \big(-48+52 \alpha -70 \alpha ^2+20 \alpha ^3-2 \alpha ^4\big) y_2+\\&&\frac{1}{3} 4^{-2+\alpha } \big(24-18 \alpha +23 \alpha ^2-6 \alpha ^3+\alpha ^4\big) y_3+\\&&
 \frac{1}{3} 4^{-2+\alpha } \big(-8 \alpha +12 \alpha ^2-4 \alpha ^3\big) y_4 \\
SRM_0&=&
 \frac{1}{3} 4^{-2+\alpha } \alpha  \big(-6+11 \alpha -6 \alpha ^2+\alpha ^3\big) y_0-\\&&\frac{1}{3} 2^{-3+2 \alpha } \big(24-46 \alpha +29 \alpha ^2-8 \alpha ^3+\alpha ^4\big) y_1+\\&&
 \frac{1}{3} 4^{-2+\alpha } \big(24-18 \alpha +23 \alpha ^2-6 \alpha ^3+\alpha ^4\big) y_2+\\&&\frac{1}{3} 4^{-2+\alpha } \big(-56 \alpha +12 \alpha ^2-4 \alpha ^3\big) y_3+\\&&
 \frac{1}{3} 4^{-2+\alpha } \big(24-12 \alpha +12 \alpha ^2\big) y_4 \\
SRM_0&=&
 \frac{1}{3} 2^{-3+2 \alpha } \big(-24+50 \alpha -35 \alpha ^2+10 \alpha ^3-\alpha ^4\big) y_0+\\&&\frac{1}{3} 2^{-3+2 \alpha } \alpha  \big(-6+11 \alpha -6 \alpha ^2+\alpha ^3\big) y_1\\&&
 -\frac{1}{3} 2^{-1+2 \alpha } \alpha  \big(2-3 \alpha +\alpha ^2\big) y_2+\\&&2^{-1+2 \alpha } \big(2-\alpha +\alpha ^2\big) y_3\\&&
 -2^{2 \alpha } \alpha  y_4
\end{array}
\]

One can easily check by substitution, that when $y_0=y_1=y_2=y_3=y_4$, each derivative is zero $SRM_0=SRM_1=SRM_2=SRM_3=SRM_4=0$, and when
$y_0=1, y_1=0.5, y_2=0,y_3=-0.5,y_4=-1$, the sum is zero $SRM_0+SRM_1+SRM_2+SRM_3+SRM_4\;=\;0$.

\section{Figure captions}
\begin{figure}[ht]
\begin{center}
\includegraphics[width=0.85\textwidth]{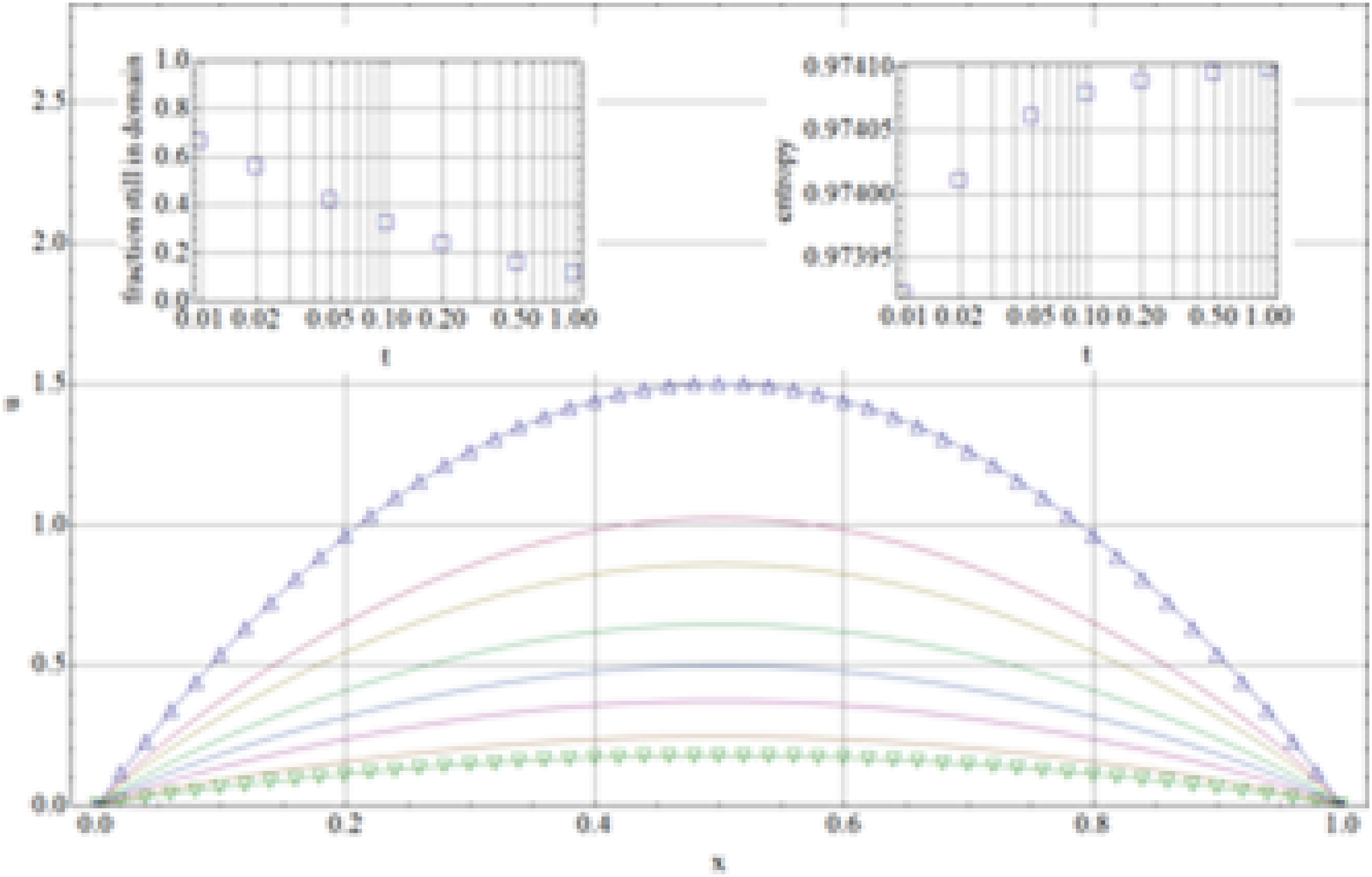}
\end{center}
\caption{
Cauchy problem with Dirichlet boundary conditions. Results from matrix method $\{\beta=1/2,\;\alpha=3/,\;a=1,\; a'=\sqrt{2}\}$. Upper triangles: initial condition, $f_i(x )=6 x(1-x),\; 0 \leq x \leq 1$. Lower triangles: solution at time $t=1$. The solid lines show density distributions at intermediate times $\{0.01,0.02,0.05,0.1,0.2,0.5\}$. The inserts show the fraction of substance still in the domain and the evolution of normalized entropy (see Appendix).
}
\end{figure}

\begin{figure}[ht]
\begin{center}
\includegraphics[width=0.85\textwidth]{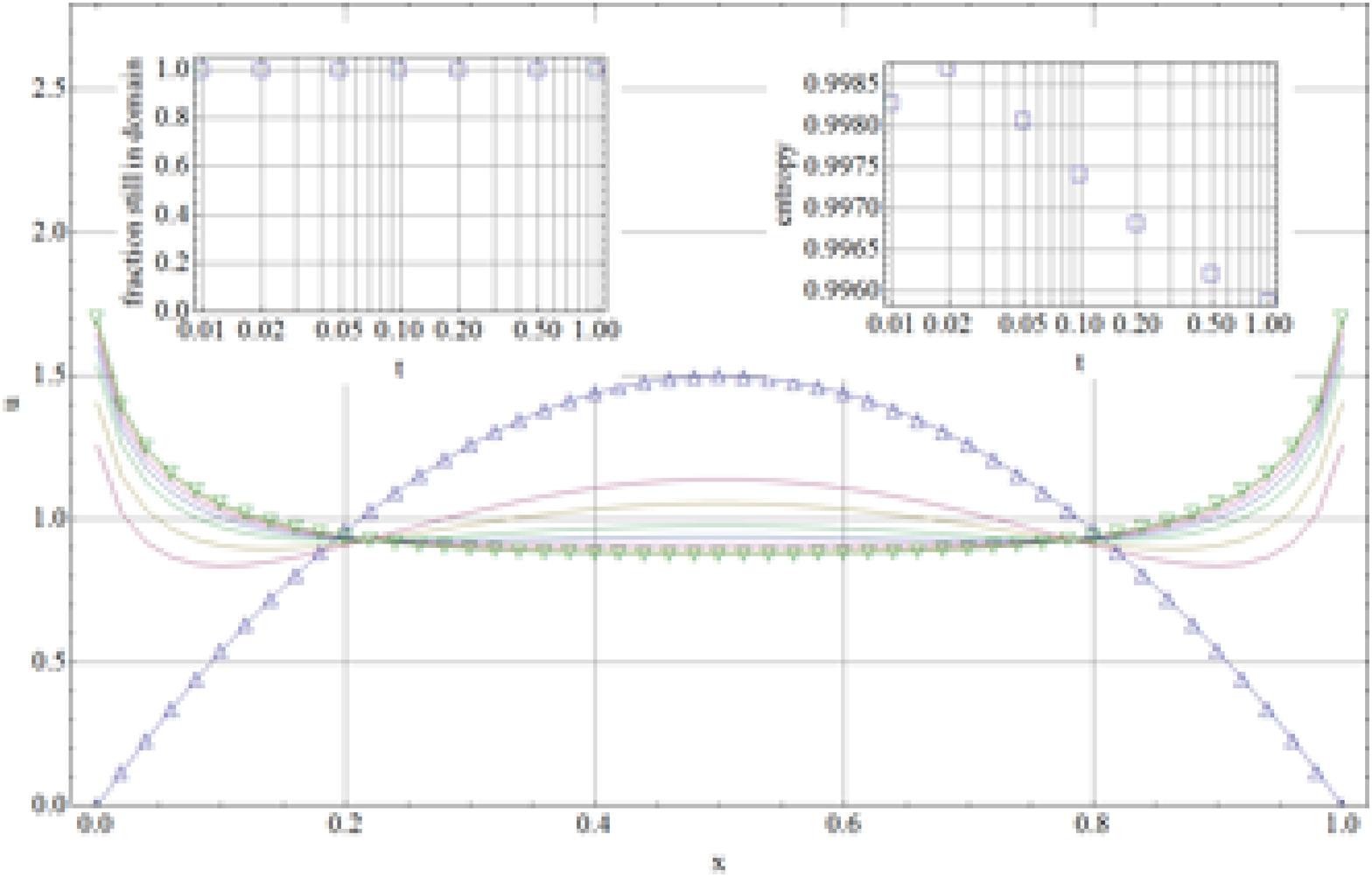}
\end{center}
\caption{
Cauchy problem with fixed amount of substance and symmetry condition. Results from Matrix method $\{\beta=1/2,\;\alpha=3/2,\;a=1,\; a'=\sqrt{2}\}$ . Upper triangles: initial condition, $f_i(x )=6 x(1-x),\; 0 \leq x \leq 1$. Lower triangles: solution at time $t=1$. The solid lines show density distributions at intermediate times $\{0.01,0.02,0.05,0.1,0.2,0.5\}$. The inserts show the fraction of substance still in the domain and the evolution of normalized entropy.
}
\end{figure}

\begin{figure}[ht]
\begin{center}
\includegraphics[width=0.85\textwidth]{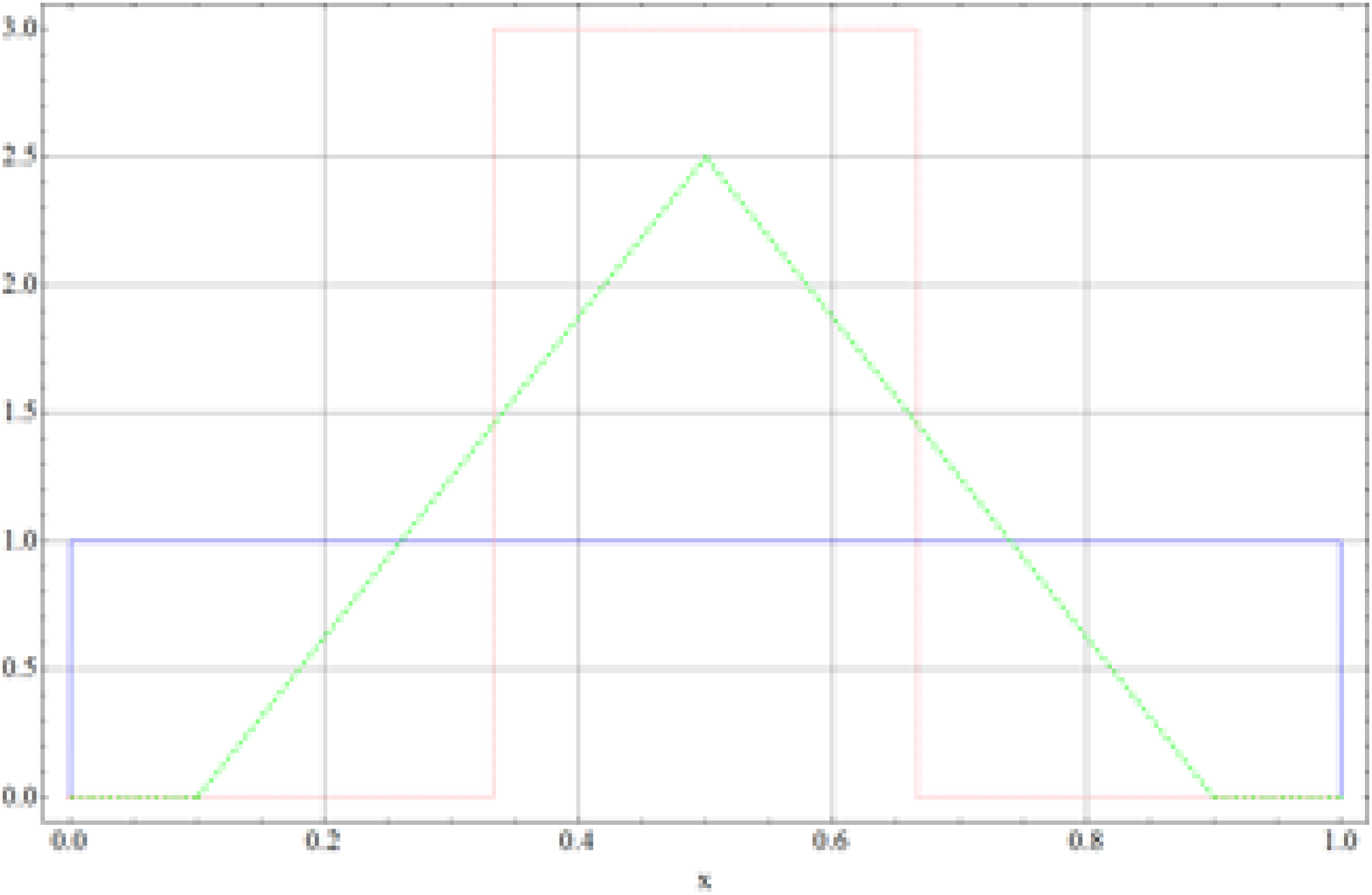}
\end{center}
\caption{
a)  Initial condition uniform distribution (solid), b) triangular distribution (dashed), c) narrow uniform distribution (dotted). Prescription (13) will  either contradict first or second law for  a).  Prescription (14) will contradict second law for b) and c). Prescription (15) will contradict second law for c).
}
\end{figure}
 	
\begin{figure}[ht]
\includegraphics[width=0.5\textwidth]{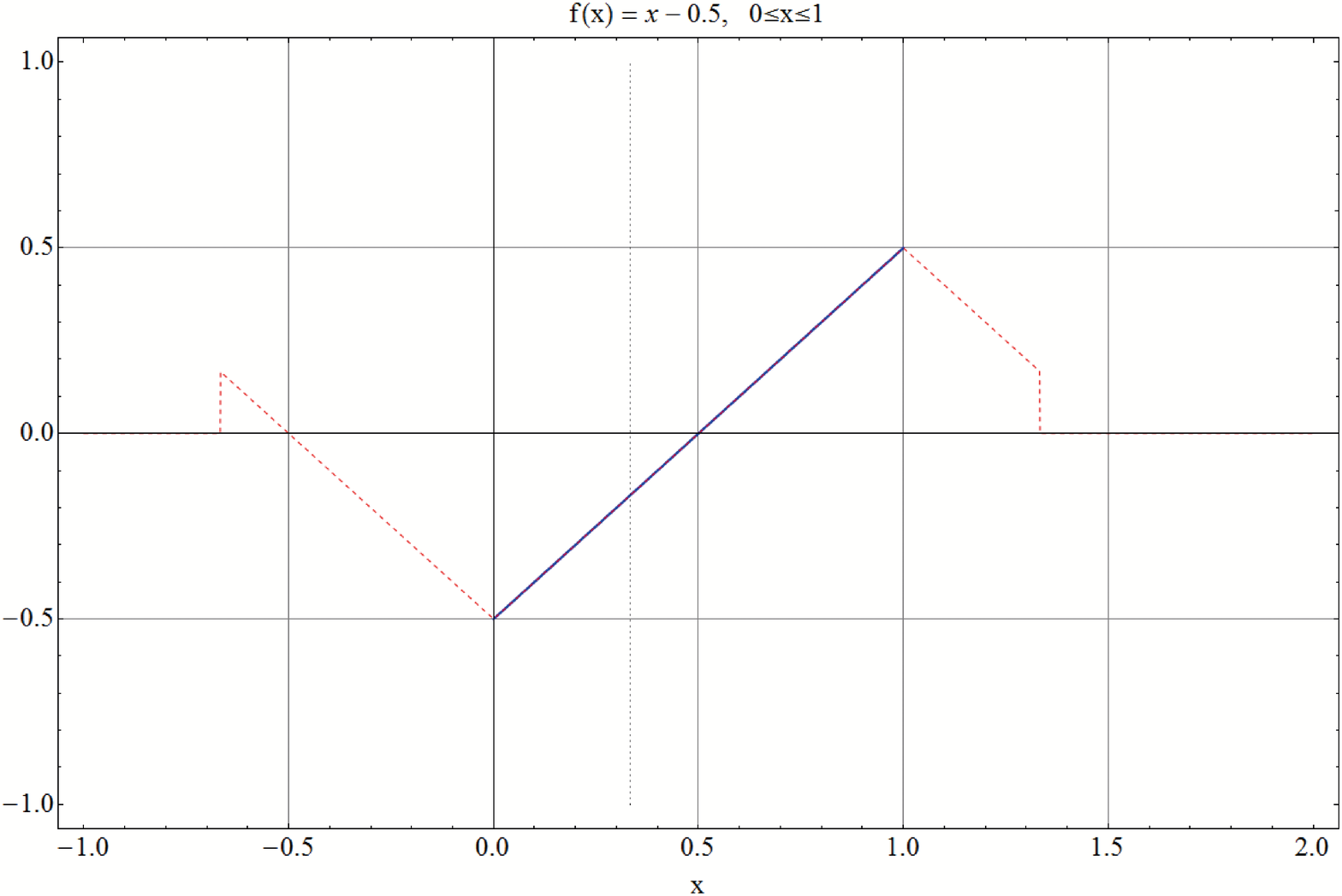}
\includegraphics[width=0.5\textwidth]{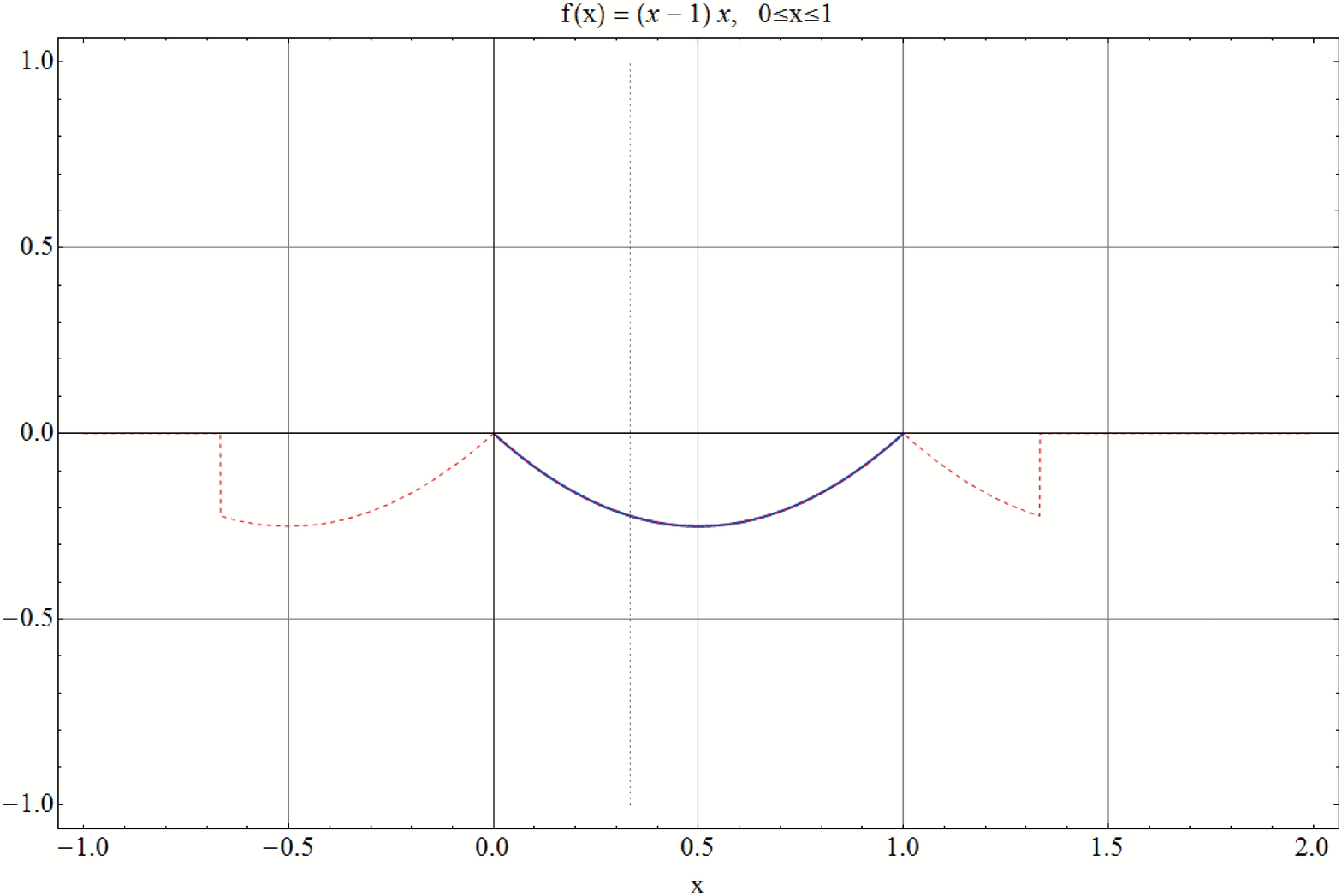}
\caption{
Illustration to prescription (17). Construction of $f_l$ and $f_r$ (dashed) from $f(x)$ given over  $0 \leq x \leq 1$ (solid) for a specified $x=1/3$ (dotted). Dashed line, left from $x=1/3$ corresponds to $f_l$, right from $x=1/3$ to $f_r$.
}
\end{figure}

\begin{figure}[ht]
\includegraphics[width=0.5\textwidth]{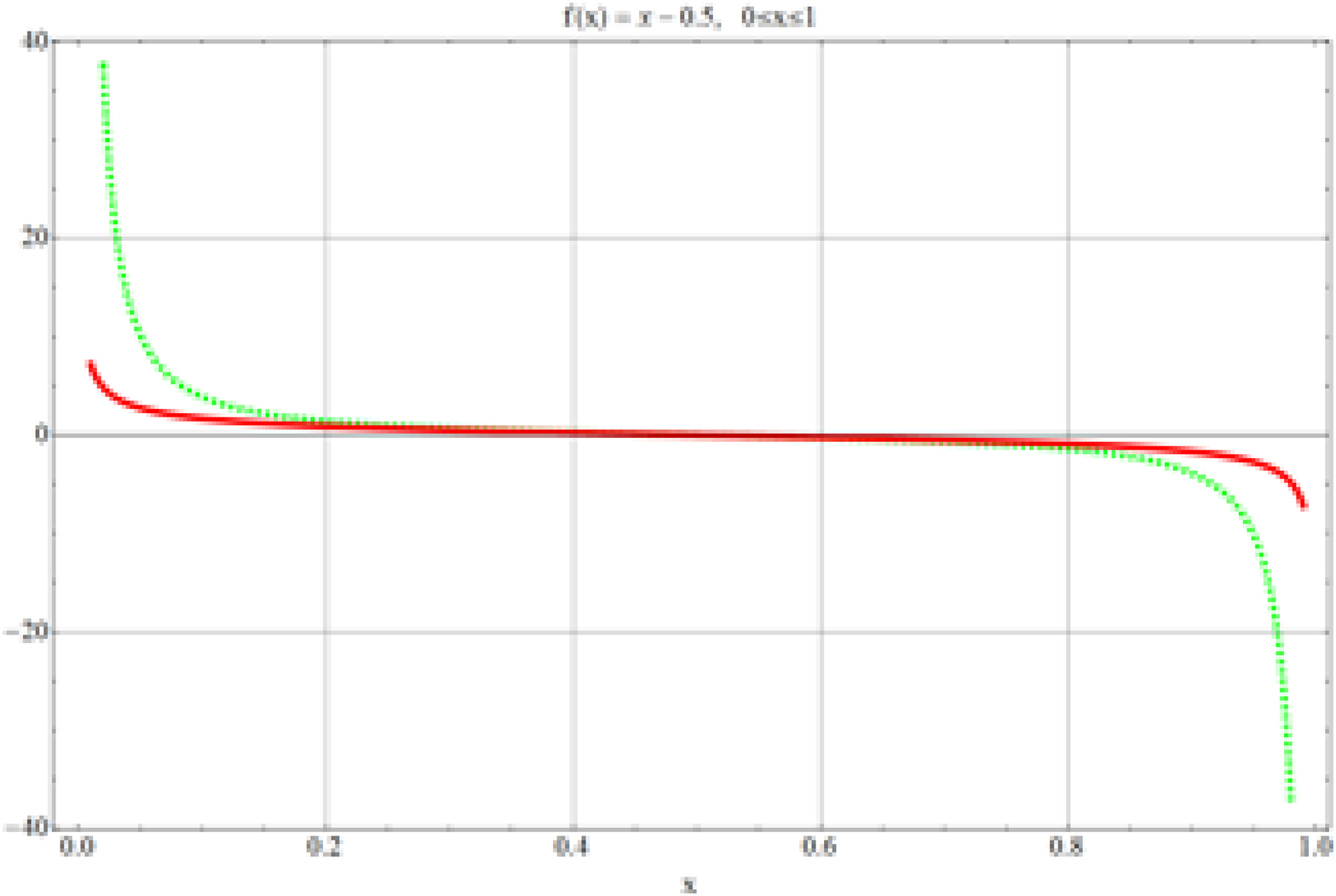}
\includegraphics[width=0.5\textwidth]{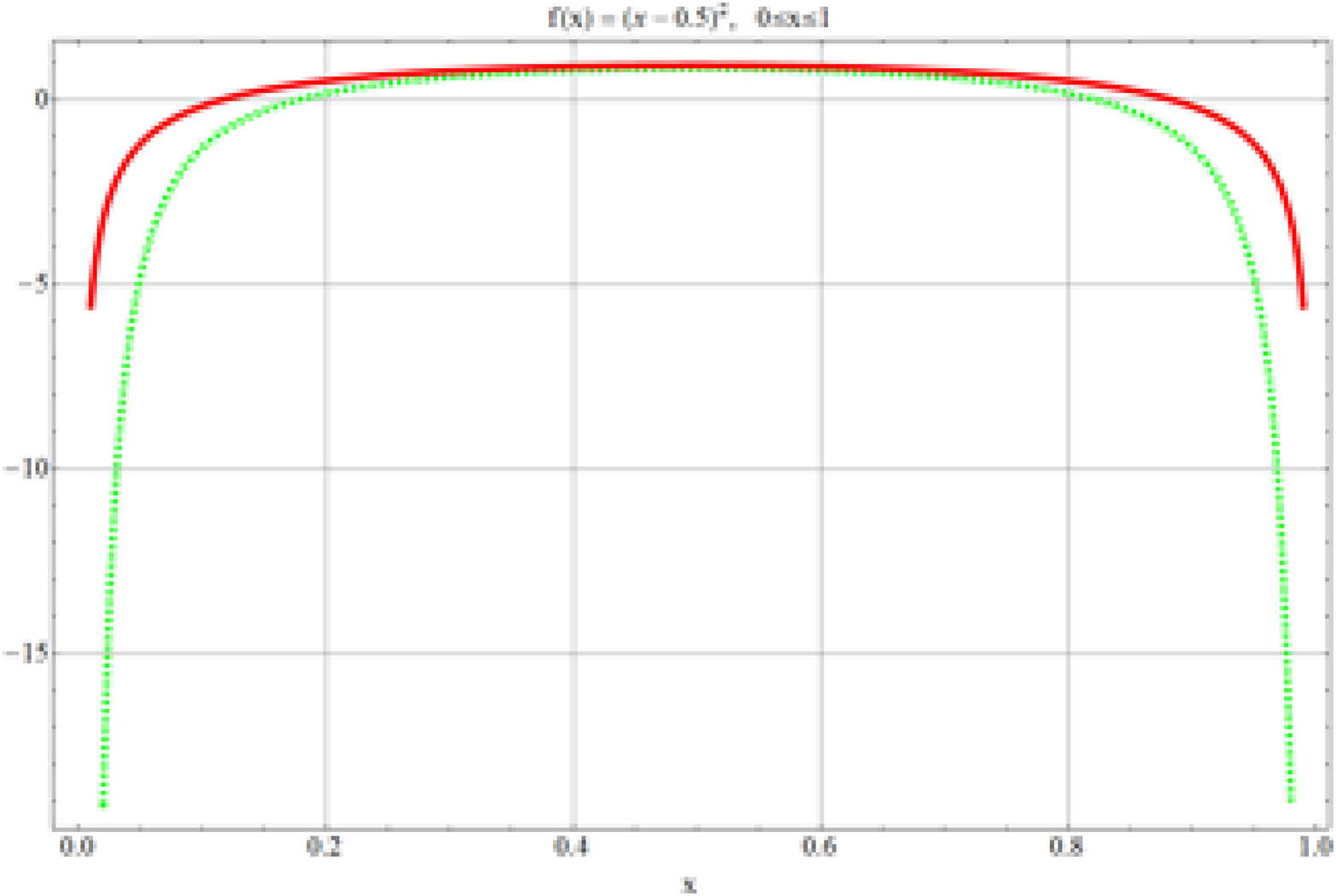}
\includegraphics[width=0.5\textwidth]{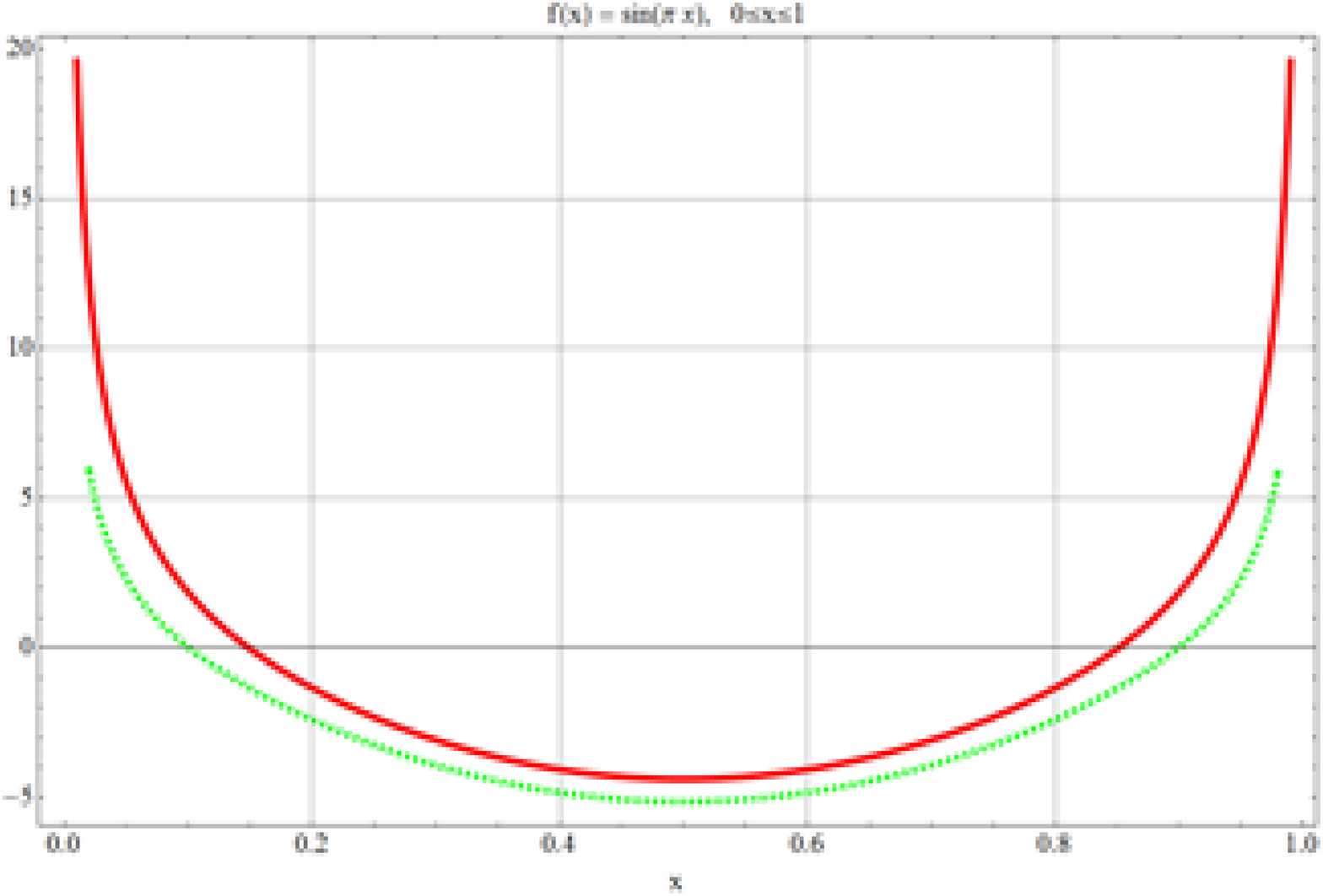}
\includegraphics[width=0.5\textwidth]{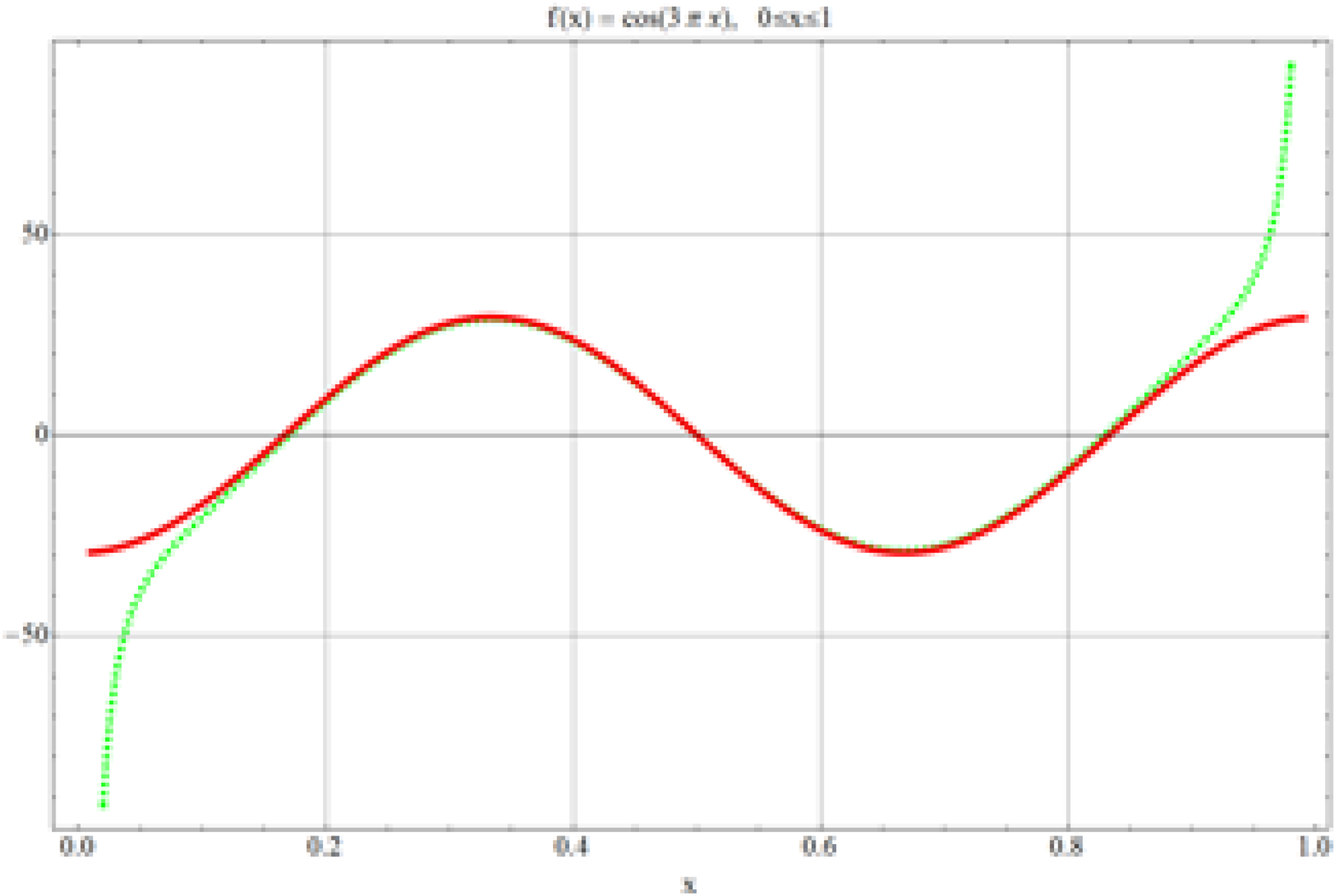}
\caption{
Modified finite Riesz derivative -- prescription (17)  (solid) and standard Riesz derivative -- prescription (13) (dashed) for various $f(x),\;  0\leq x \leq 1$ functions when $\alpha = 3/2$.  For the cosine function we also show the Fourier-Riesz derivative of the unbounded domain function -- Equation (6) (dotted); it virtually coincides with the modified Riesz derivative - prescription (17), apart from a factor of $1.01328\ldots$,  see Equation (24).
}
\end{figure}

\begin{figure}[ht]
\begin{center}
\includegraphics[width=0.85\textwidth]{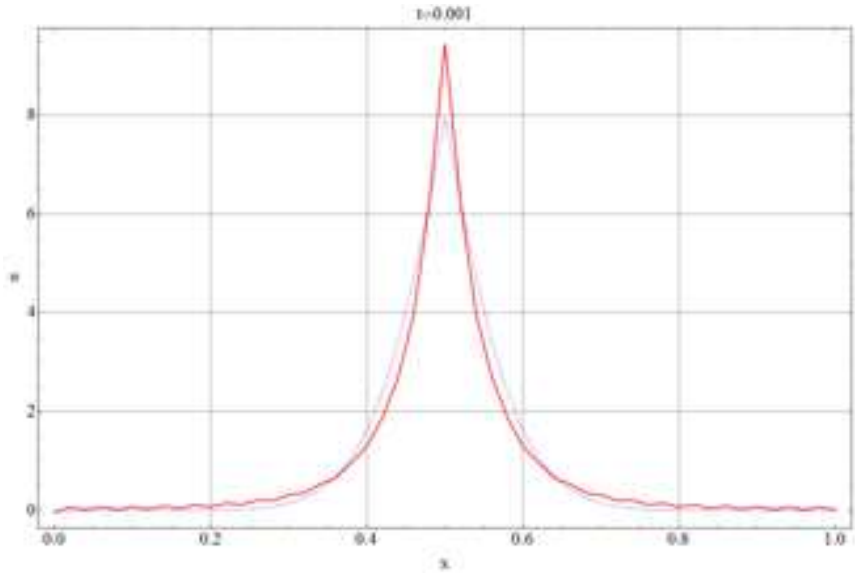}
\end{center}
\caption{
Fundamental solution of the bounded domain problem, Equation (36) with 50 terms (solid) and fundamental solution of the unbounded problem Equation (37) at an early time $t=0.001$ for $\{\beta=1/2,\;\alpha=3/2,\;a=1\}$.
}
\end{figure}

\begin{figure}[ht]
\begin{center}
\includegraphics[width=0.85\textwidth]{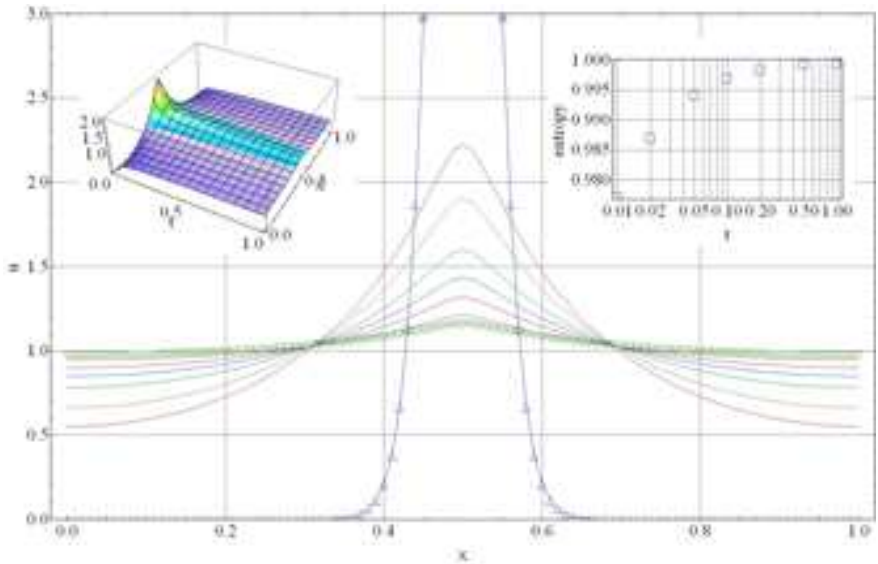}
\end{center}
\caption{
Fundamental solution of the bounded domain problem $\{\beta=1/2,\;\alpha=3/2,\;a=1\}$ calculated from the LT-collocation method with numerical inversion. Upper triangle: initial condition, taken as Equation (37) at a very early time $t=0.00001$. Lower triangle, solution at time $t=1$. The solid lines show density distributions at intermediate times $\{0.01,0.02,0.05,0.1,0.2,0.5\}$.  One insert shows the 3D surface and the other the evolution of normalized entropy.
}
\end{figure}

\begin{figure}[ht]
\begin{center}
\includegraphics[width=0.85\textwidth]{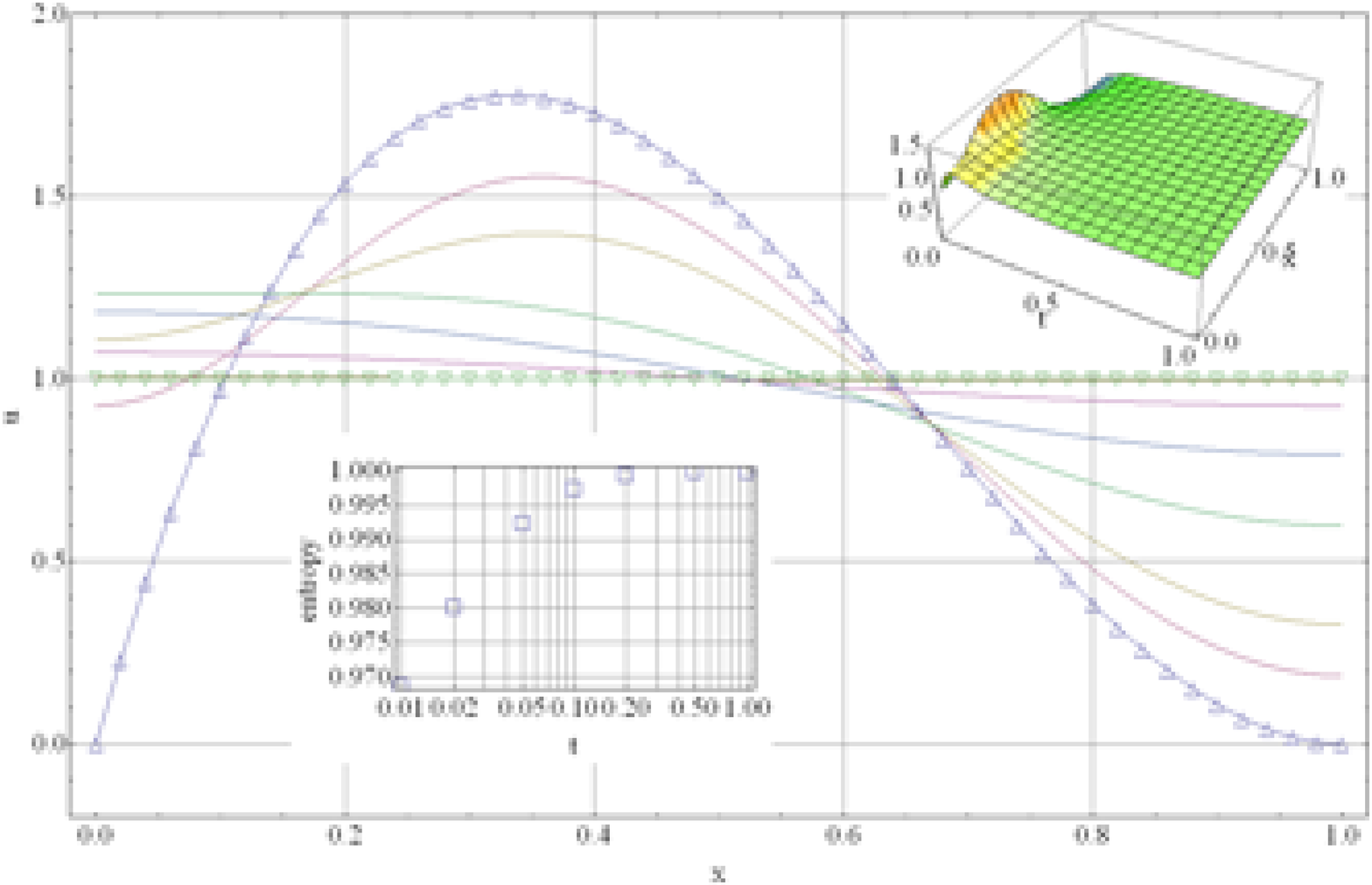}
\end{center}
\caption{
Results from LT-collocation method with numerical inversion $\{\beta=1,\;\alpha=2,\;a=1\}$  coincide with known results. Upper triangles: initial condition, and $f_i(x )=12 x(1-x)^2,\;0\leq x \leq 1$. Lower triangles: solution at time $t=1$. The solid lines show density distributions at intermediate times $\{0.01,0.02,0.05,0.1,0.2,0.5\}$. The inserts show the 3D surface and the evolution of normalized entropy.
}
\end{figure}

\begin{figure}[ht]
\begin{center}
\includegraphics[width=0.85\textwidth]{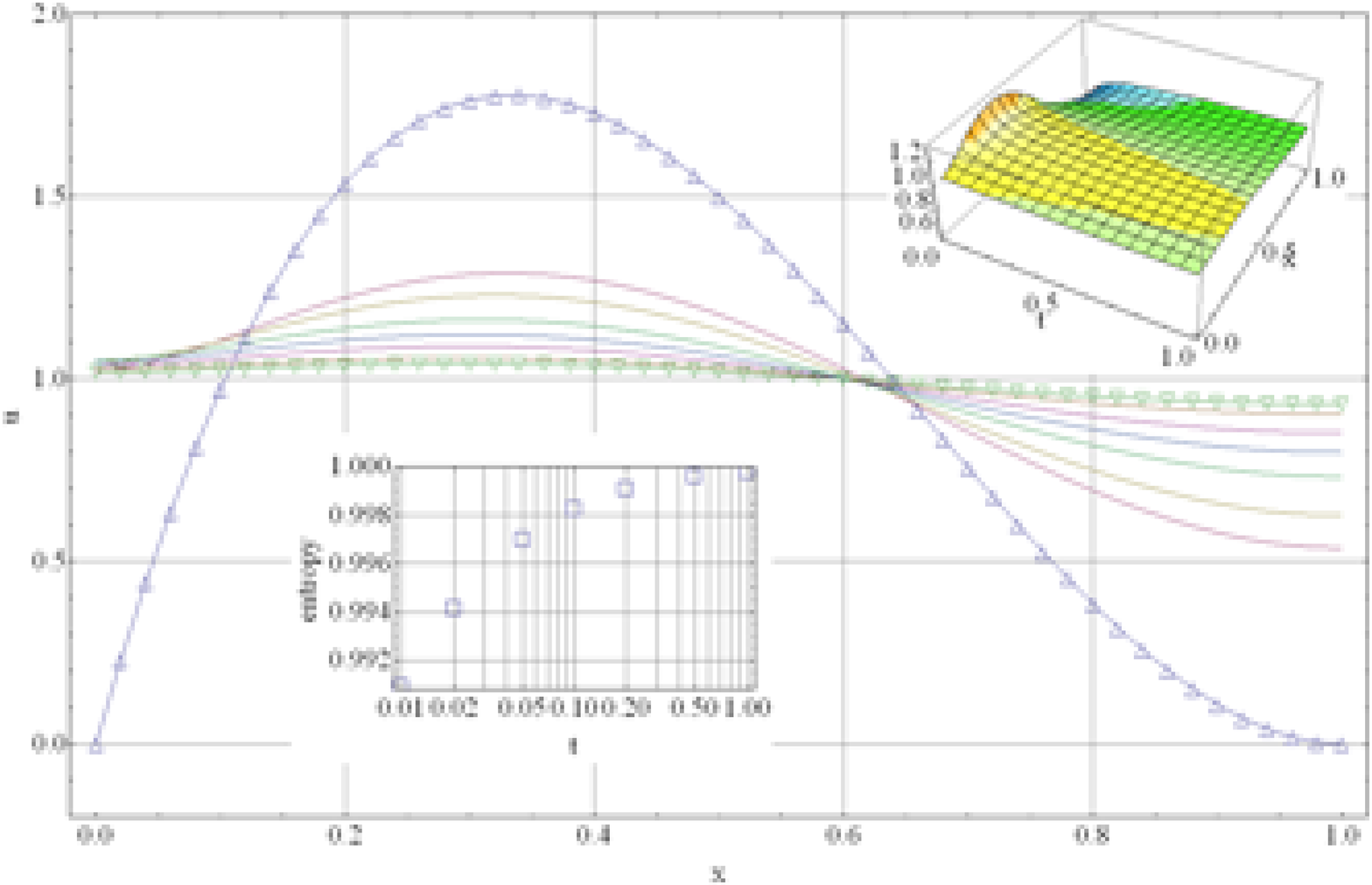}
\end{center}
\caption{
Results from LT-collocation method with numerical inversion for $\{\beta=1/2,\;\alpha=3/2,\;a=1\}$. Upper triangles: initial condition $f_i(x )=12 x(1-x)^2,\; 0\leq x \leq 1$. Lower triangles: solution at time $t=1$. The solid lines show density distributions at intermediate times $\{0.01,0.02,0.05,0.1,0.2,0.5\}$. The inserts show the 3D surface and the evolution of normalized entropy. In contrast to Fig. 8, there is less "overshoot" at $x = 0$ and the entropy values are higher, except for $t = 1$.
}
\end{figure}

\begin{figure}[ht]
\begin{center}
\includegraphics[width=0.85\textwidth]{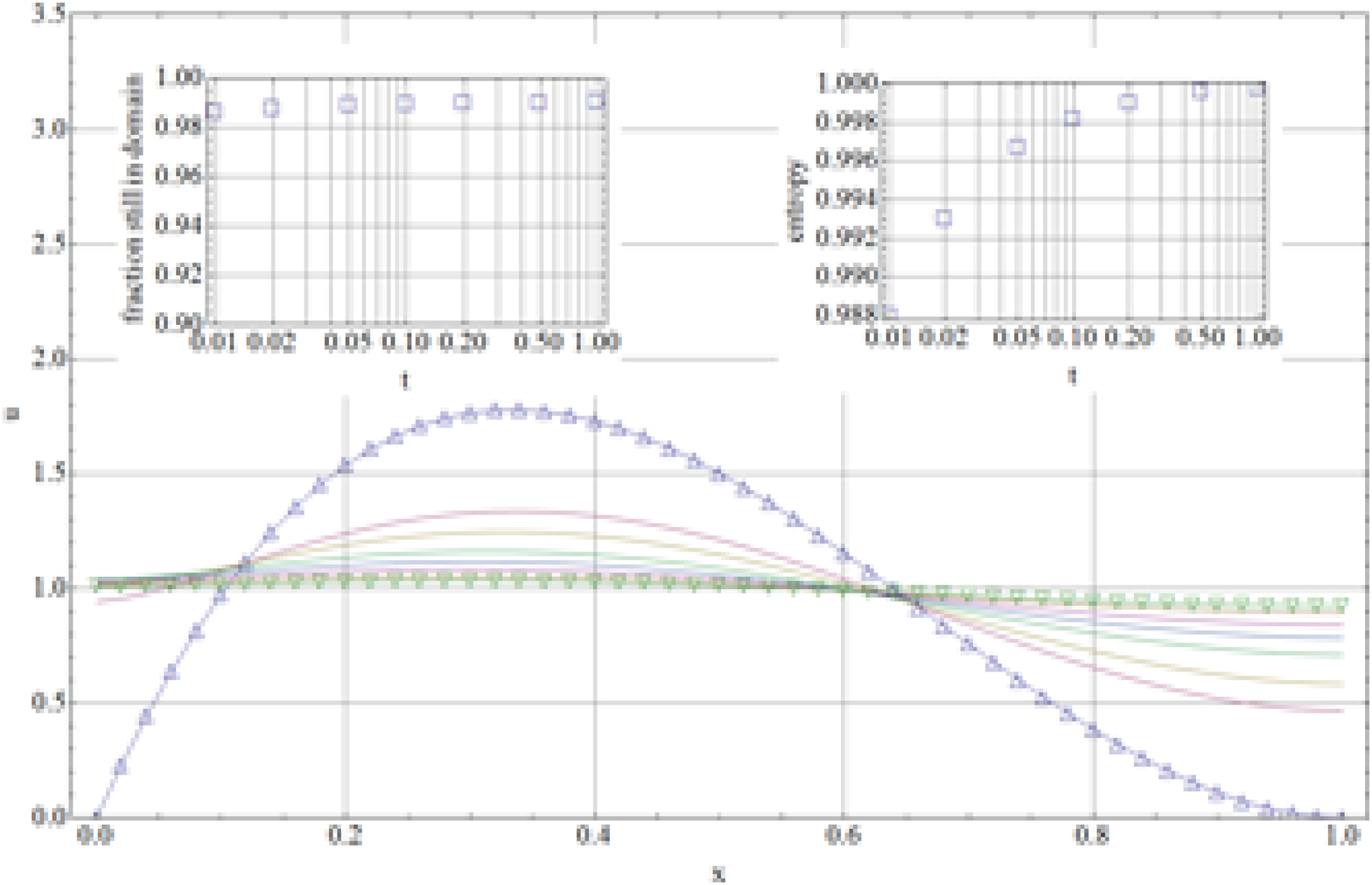}
\end{center}
\caption{
Results from (extended) the matrix approach using the modified discrete operator  (see Appendix) without any specific  boundary conditions for $\{\beta=1/2,\;\alpha=3/2,\;a=1,\; a'=\sqrt{2}\}$. Upper triangles: initial condition $f_i(x )=12 x(1-x)^2,\; 0 \leq x \leq 1$. Lower triangles: solution at time $t=1$. The solid lines show density distributions at intermediate times $\{0.01,0.02,0.05,0.1,0.2,0.5\}$. The inserts show the fraction of substance still in the domain and the evolution of normalized entropy. Notice the slight deviation from Fig. 9, because of the limited accuracy of the finite difference approach with $\Delta{x}=0.02$ and $\Delta{t}=0.01$.
}
\end{figure}

\end{document}